\documentclass[aps,prb,twocolumn,longbibliography,groupedaddress]{revtex4-2}%
\usepackage{amssymb}
\usepackage{amsfonts}
\usepackage{amsmath}
\usepackage{graphicx}%
\providecommand{\U}[1]{\protect\rule{.1in}{.1in}}
\begin{document}
\title{Altermagnetism in MnTe: origin, predicted manifestations, and routes to detwinning}
\author{I. I. Mazin}
\affiliation{Department of Physics and Astronomy, George Mason University, Fairfax, VA
22030, USA}
\affiliation{Quantum Science and Engineering Center, George Mason University, Fairfax, VA
22030, USA}

\begin{abstract}
MnTe has recently attracted attention as an altermagnetic candidate.
Experimentally it has an altermagnetic order of ferromagnetic $ab$ planes,
stacked antiferromagnetically along $c$. We show that this magnetic order (by
itself non-trivial, since the in-plane exchange in antiferromagnetic) opens
intriguing possibility of manufacturing altermagnetically-detwinned samples
and generate observable magnetooptical response (which we calculate from first
principles) as a signature of altermagnetism.

\end{abstract}
\maketitle

The recently discovered phenomenon of spin-split bands in collinear
symmetry-compensated antiferromagnets, dubbed \textquotedblleft
altermagnetism\textquotedblright\ (AM)\cite{PRX,Punch,perspective}, has
attracted considerable attention. While a number of altermagnets have been
theoretically identified, there is a big experimental challenges in assessing
this, for a number of reasons: First, most of them are not metals, so
anomalous Hall conductivity cannot be measured. Second, many have the easy
magnetization direction not compatible with anomalous response. Third,
statistically these materials form chiral domains, so that the anomalous
response of opposite signs largely cancels.

There are ways to overcome these difficulties. First, since the nondiagonal
optical conductivity, accessible through magnetooptical effects, is governed
by the same selection rules as the anomalous Hall conductivity, it can be used
in its place to detect the AM response. An additional advantage is that, as
discussed later in the paper, calculations of the finite-frequency response
from the first principles is mush easier and more reliable that in the static
(Hall) limit. Finally, while the chiral domains necessarily form
statistically, as the magnetic phase is nucleating upon cooling simultaneously
in different parts of the sample, it does not carry, as opposed to
ferromagnetics, any energetic advantage, only the energy cost of forming
domain walls. This suggests that carefull annealing through the Neel
temperature, preferably with a temperature gradient, in order to suppress
independent nucleation in different parts of the sample, or on a ferromagnetic
substrate, in order to encourage a single domain on the interface, may result
in a single domain sample, or domains large enough to be probed by polarized
light independently. However, before urging experimentalists to pursue this
path, a better and more quantitative understanding of this material is imperative.

Specifically, two main issues need to be understood: (i) magnetic interactions
in MnTe, as they eventually determing the domain wall dynamics, and (ii)
frequencies at which the strongest magnetoptical response is expected, and an
estimate of the latter. In this paper we will provide both.

MnTe crystallizes in the NiAs crystal structure, as is known since
1956\cite{hex}, which can be viewed as the hexagonal analog of the metastable
cubic MnTe (crystallized in the NaCl structure)\cite{cubic}. In the latter,
both Mn and O form triangular layers stacked along 111 as AbCaBc (the
uppercase letters correspond to the Mn layers). In the former, the stacking
sequence is AbAc, and the structure is expanded in the direction perpendicular
to the triangular planes, and squeezed in the planes (Fig. 1).

As a result, while the Mn-Mn interlayer distance is 2.60 \AA \ in the cubuc
MnTe, it is 3.37 \AA \ in the hexagonal one, which is also the shortes Mn-Mn
bond. The next bond connects two Mn in the $ab$ plane, and is 4.15 \AA \ long;
both are shorter than the corresponding bonds in the cubic material, which is
4.23 \AA . The corresponding Mn-Te-Mn angles (Fig. 2) are 70.3$%
{{}^\circ}%
$ and 90.1$%
{{}^\circ}%
.$ The third neighbors correspond to the second neighbors in the cubic
structure, where they are bridged by Te along the straight line (a 180$%
{{}^\circ}%
$ angles) and the distance is 5.98 \AA ; in the hexagonal structure it is 5.35
\AA \ and the angle is 131.7$%
{{}^\circ}%
$.

MnTe has been studied a lot, both experimentally and theoretically. The latest
and the most comprehensive study was probably Ref. \cite{Prague} (see also the
references therein). Experimentally, there is full consensus that MnTe forms
an A-type antiferromagnetic structure with $\mathbf{q}=(0,0,0),$ and the
magnetic moments are collinear and aligned with the (210) direction (i.e.,
perpendicular to the Mn-Mn bond). The in-plane magnetic anisotropy energy $K$
was found to be too small to be measured by neutrons in Ref. \cite{MnTe-SW},
and too small to be calculated reliably in Ref. \cite{Prague}. The in-plane
spin-flop field in Ref. \cite{Prague} was between 2 and 6 T, which, using the
leading exchange coupling of $J\sim40$ meV (see below), corresponds to
$K\approx\sqrt{2KJ}\approx0.2-1.4$ $\mu$eV.

Spin-wave dispersion was fitted with three nearest neighbor Heisenberg
exchange coupling, defined via the Hamiltonian%
\begin{equation}
H=\sum_{i=1-3}J_{i}\mathbf{\hat{m}\cdot\hat{m}}^{\prime},
\end{equation}
where the summation is over all different bonds of a given length, and
$\mathbf{\hat{m}}$, $\mathbf{\hat{m}}^{\prime}$ are the unit vectors of spins
forming the bond. The resulting parameters are listed in Table 1, together
with those calculated in Ref. \cite{MnTe-VASP} and our own calculations.
\begin{table}[h]
\caption{{Calculated and experimental Heisenberg exchange parameters, in
meV.}}
\begin{center}%
\begin{tabular}
[c]{|l|c|c|c|c|c|}\hline
& $J_{1}$ & $J_{2}$ & $J_{3}$ & $J_{4}$ & $T_{CW}$ (K)\\\hline
Expt. (\cite{MnTe-SW}) & 46.2 & -1.44 & 6.2 & - & 612$^{a}$, 585$^{b}$\\
Calc. (\cite{MnTe-VASP}) & 38.4 & 0.34 & 5.0 & 2.0 & 552\\
Calc. (this work) & 42.1 & 0.91 & 5.3 & - & 592\\\hline
\end{tabular}
\end{center}
\par
$^{a}$ calculated from the exchange parameters in Ref. \cite{MnTe-SW}.
\newline$^{b}$ measured\cite{TCW,TCW2}.\label{T1}%
\end{table}Note that both DFT calculations, while performed by different
methods (VASP\cite{PAW2} in Ref. \cite{MnTe-VASP}, LAPW\cite{Wien2k} here),
give the nearest-neighbor in-plane exchange $J_{2}$ antiferromagnetic, while
Ref. \cite{MnTe-SW} reports a very small ferromagnetic value. We believe that
this is an experimental artifact, maybe due to neglect of the longer
interactions in the spin-wave analysis. Indeed, for Mn$^{2+}$ there is no
superexchange mechanism that could generate a ferromagnetic coupling, and no
itinerant electrons to promote ferromagnetism. Since the bond angle in this
case is nearly exactly 90$%
{{}^\circ}%
,$ only $pd\sigma\times pd\pi$ superexchange processes are allowed, but, since
both $t_{2g}$ and $e_{g}$ states are occupied, their contribution is
antiferromagnetic (as opposed to, for instance, Cr$^{3+}),$ and proportional
to $t_{pd\sigma}^{2}t_{pd\pi}^{2}/U\Delta^{2},$ where $U$ is the Hubbard
repulsion and $\Delta$ is the Mn($d)-$Te($\pi)$ energy separation. The
Goodenough-Kanamori ferromagnetic exchange is of course present, but
proportional to $J_{H}($O$)(t_{pd\sigma}^{4}+t_{pd\pi}^{4})/\Delta^{2},$ which
is much smaller.

With this in mind, one way wonder what drives the ferromagnetic order in
plane. The answer is that this is $J_{3},$ which is sizable and has high
degeneracy of 12, and tries to make the nearest neighbors in the plane
antiparallel to the once-removed Mn in the neigboring plane, that is, parallel
to each other. It can thus easily overcome the antiferromagnetic $J_{2}.$

\begin{figure}[th]\includegraphics[width=.35\linewidth]{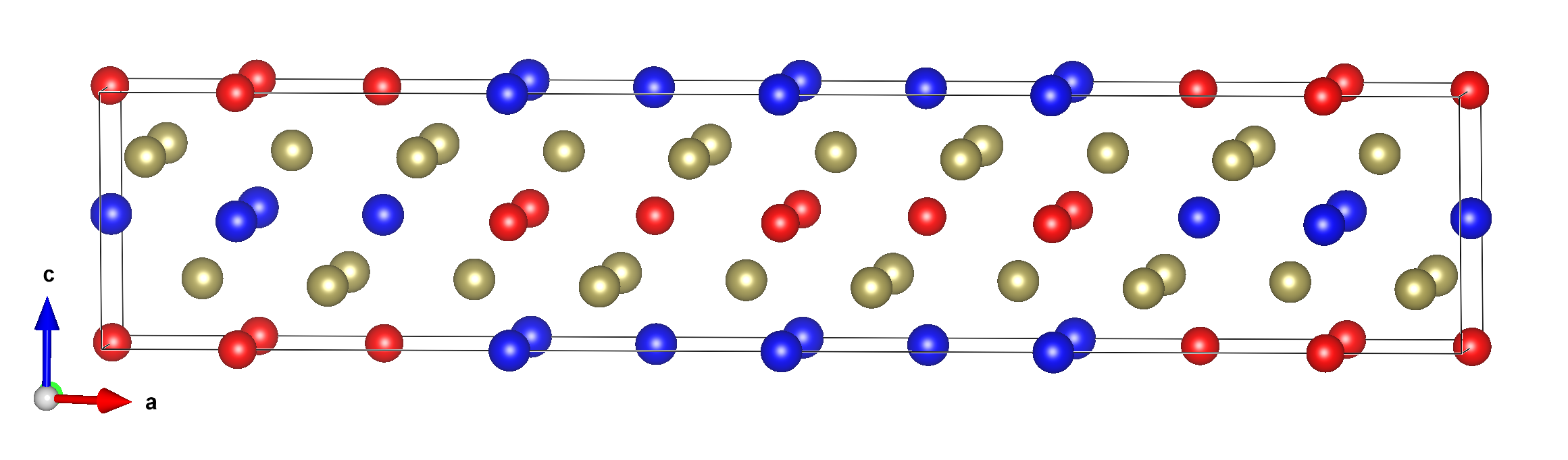}\includegraphics[width=.64\linewidth]{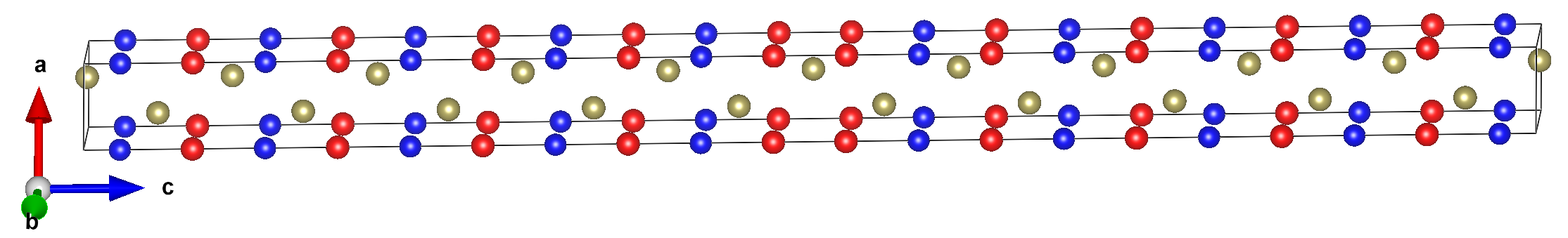}
\caption{Supercells used for the domain wall energy calculations for an $ab$ domain (left)
and a $c$ domain (right).}%
\label{domains}%
\end{figure}

These findings suggest that the $ab$ domain walls, that is to say, walls
perperdicular to the $ab$ plane, should form more easily that those parallel
to $ab$ (see Fig. \ref{domains} We have verified that through direct density
functional (DFT) calculations, using the standard VASP package\cite{PAW2},
with the following settings: a 20 formula units supercell, the k-point mesh
parallel to the domain boundary 12x12, perpendicular 3, pseudopotentials
PAW\_PBE Te and Mn\_pv, energy cutoff 400 eV, and applying $U-J=4$ eV, which
gives a reasonable direct optical gap of 1.7 eV and indirect gap of 0.8 eV.
The results are shown in Table\ref{T2}, where we also show the effect of
lattice optimization (positions only). \begin{table}[h]
\caption{Calculated energy of the domain walls, in meV per Mn at boundary.}%
\label{T2}
\begin{center}%
\begin{tabular}
[c]{|c|c|c|c|}\hline
\multicolumn{2}{|c|}{$ab$ domain} & \multicolumn{2}{|c|}{$c$ domain}\\\hline
not optmized & optimized & not optmized & optimized\\\hline
19.1 & 19.0 & 65.2 & 55.4\\\hline
\end{tabular}
\end{center}
\end{table}

As expected, the $c$ wall has a much higher energy and is much less
likely to form. On the other hand, since individual $ab$ planes are
ferromagnetic, growing MnTe on a single-domain ferromagnetic substrate (with
can be easily achieved by applying an in-plane magnetic field) should prevent
the $ab$ domains from forming. Numerous antiferromagnets and ferromagnets with
stacked ferromagnetic layer with an in-layer easy axis are know, and many have
transition temperature above that of MnTe ($\sim310$ K), such as NaOsO$_{3}$
(610 K), (Sc,Ga)FeO$_{3}$ (up to 408 K), Fe$_{2}$O$_{3}$ (960 K), Mn$_{3}%
$(Cu,Ge) (380 K), FeBO$_{3}$ (348 K), CuMnAs (480 K), but especially promising
is LiMn$_{6}$Sn$_{6}$, which in naturally layered, has $T_{C}\approx380$ K,
and, in addition, has a nearly perfect epitaxial match with MnTe (assuming a
$\sqrt{5}\times\sqrt{5}$ superlattice, $\tilde{a}=$10.977 \AA \ for the latter
and 2$\times2,$ $\tilde{a}=$10.982 \AA \ for the former, a 0.05\% match).
While epitaxial coherence is not required, it would serve to reduce the
distance from the substrate and enhance coupling.

Thus, MnTe is a prime candidate to singe-domain altermagnetism. Unfortunately,
it is an insulator, so direct measurement of the anomalous Hall effect is not
possible. Fortunately, the altermagnetism there can be probed by
magnetooptical tools, such as MOKE (magnetooptical Kerr effect). Also
fortunately, the nondiagonal part of the optical conductivity $\sigma
_{xy}(\omega)$ can be reliably calculated by modern DFT codes, such as VASP
--- as opposed to the Hall conductivity, the zero-frequency limit of
$\sigma_{xy}(\omega),$ which is impossible to converge in existing
calculations, and all current first principle calculations rely upon
Wannier-based interpolation, which adds considerable ambiguity.
 \begin{figure}[th]
\begin{center}
\includegraphics[width=.9\linewidth]{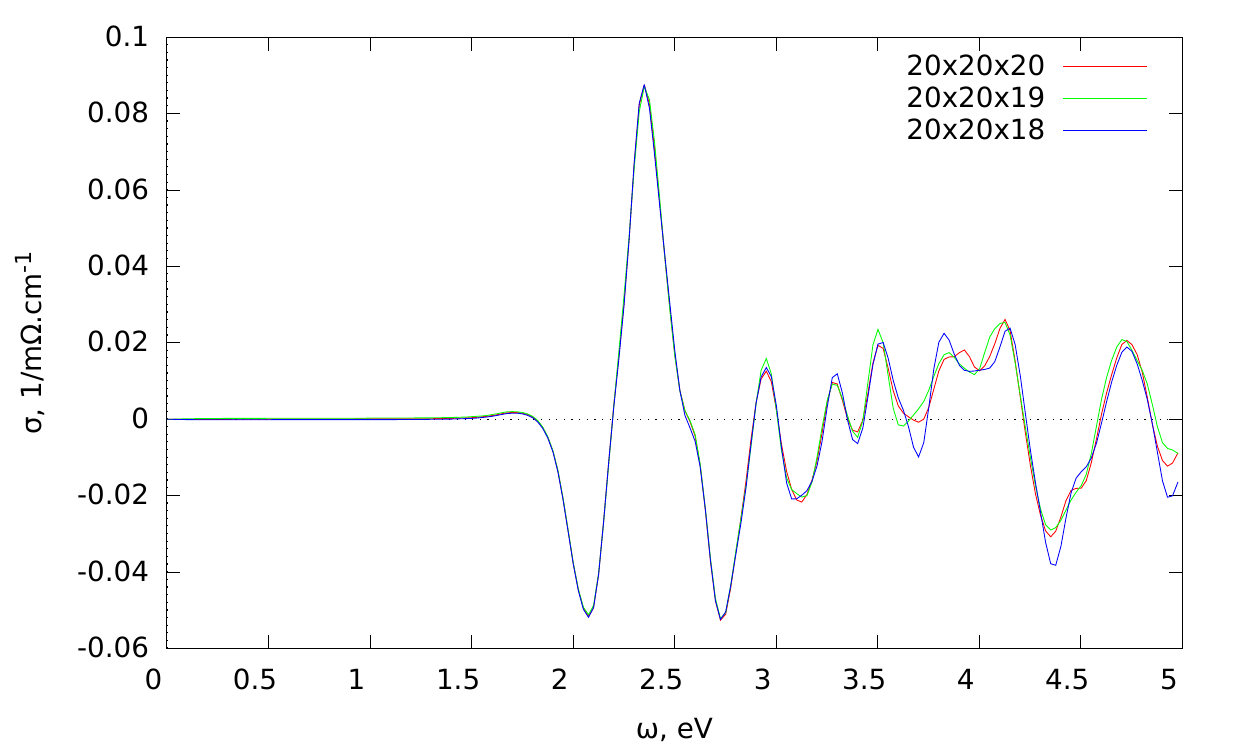}
\includegraphics[width=.9\linewidth]{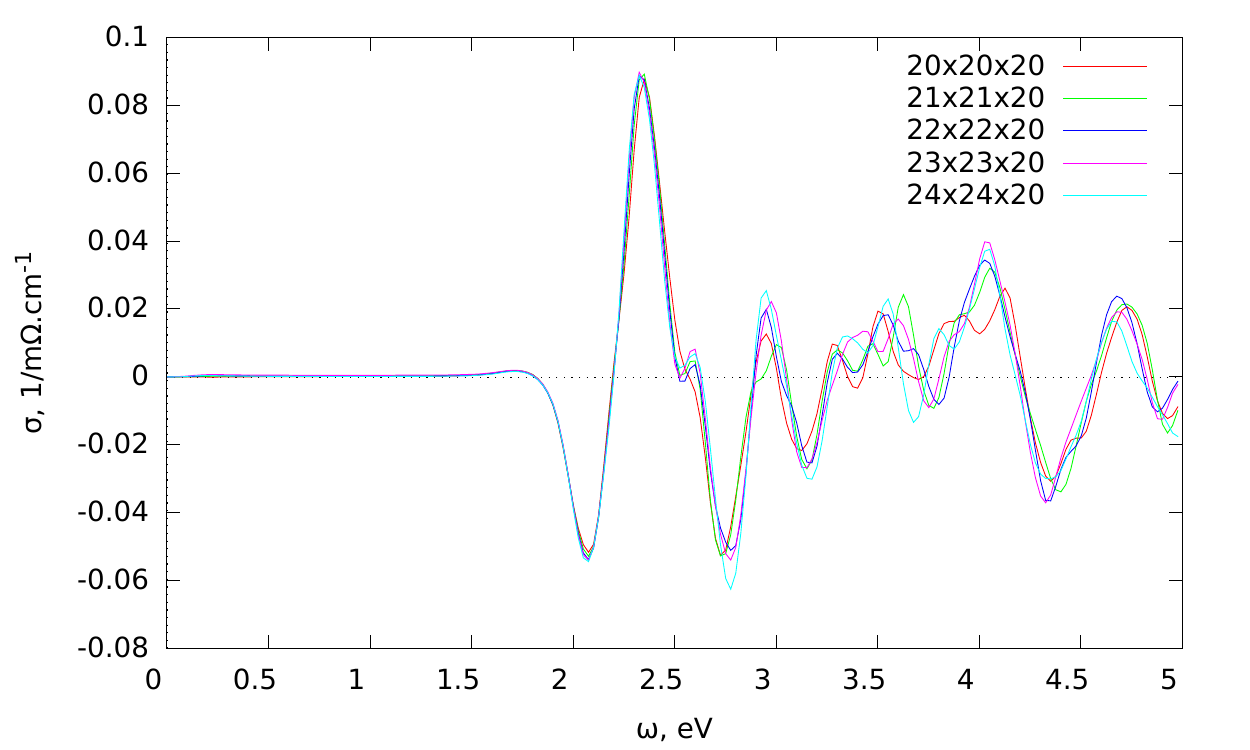}
\end{center}
\caption{Calculated nondiagonal optical conductivity $\sigma_{xy}$. The two
panels show convergence with the respect to the in-plane and out-of-plane
k-point mesh, respectively.}%
\label{conv}%
\end{figure}
In order to inform the experiments, which, we hope, will be encouraged by this
paper, we have calculated the non-diagonal part of the optical conductivity,
for the experimental easy magnetization axis of 210, that is, at $\alpha=30%
{{}^\circ}%
$ to the Mn-Mn bond. As expected, only $\sigma_{xy}$ is nonzero. We show the
convergence of $\sigma_{xy}(\omega)$ in Fig. \ref{conv}. Note that the results
are reasonably well converged already at the k-mesh of $20\times20\times20;$
for the Hall conductivity $\sigma_{xy}(0)$ in similar materials an order of
magnitude larger linear density is required. Consistent with the symmetry
analysis\cite{PRX}, only $\sigma_{xy}(\omega)$ is nonzero, and only for
$\alpha\neq0.$ In Fig. \ref{angle} we show the angular dependence of
$\sigma_{xy}(\omega)/\sin3\alpha$ as a function of $\alpha.$ One can see that
lowest-order linear dependence of $\sigma_{xy}(\omega)$ on $\sin3\alpha$ holds
with a good accuracy.

In summary, we (a) explained the microscopic origin of the ferromagnetic
ordering in the $ab$ plane of MnTe, as driven not by a ferromagnetic in-plane
exchange interaction (which has in fact the antiferromagnetic sign), but by
the second-interlayer-neighbors antiferromagnetic counplig, (b) computed the
energy of the antiferromagnetic domain walls in MnTe, and showed it to be
substantial, encouraging growing single-domaun samples, where the predicted
magnetooptica response can be measured, and (c) calculated the said response
and foud it to be sizeable, with a symmetry following the theoretical
prediction. We hope that this work will encourage experiemnatl studies of
altermagnetism in this compound.

\begin{figure}[th]
\begin{center}
\includegraphics[width=.9\linewidth]{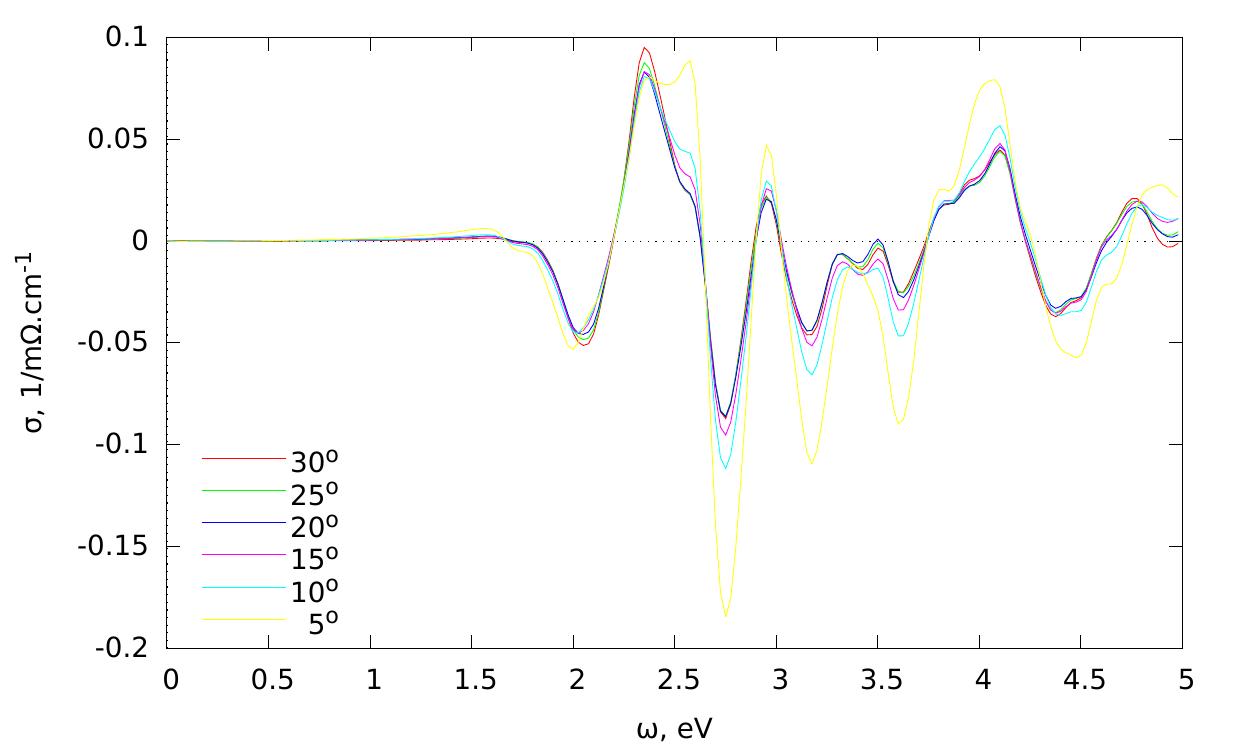}
\end{center}
\caption{Dependence of $\sigma_{xy}$ on the angle that Mn spins form with
Mn-Mn-bond direction $\alpha$ (see the inset), divided by $\sin(3\alpha)$.}%
\label{angle}%
\end{figure}\begin{acknowledgements}
The author acknowledges support from the Army Research Office
\end{acknowledgements}
\bibliographystyle{unsrt}
\bibliography{MnTe}

\end{document}